# Measuring and Visualizing Place-Based Space-Time Job Accessibility[1]

Yujie Hu[1*], Joni Downs[1]
[1]School of Geosciences, University of South Florida, Tampa, FL 33620, USA
*Corresponding author: Yujie Hu, yhu@usf.edu

**Abstract**
Place-based accessibility measures, such as the gravity-based model, are widely applied to study the spatial accessibility of workers to job opportunities in cities. However, gravity-based measures often suffer from three main limitations: (1) they are sensitive to the spatial configuration and scale of the units of analysis, which are not specifically designed for capturing job accessibility patterns and are often too coarse; (2) they omit the temporal dynamics of job opportunities and workers in the calculation, instead assuming that they remain stable over time; and (3) they do not lend themselves to dynamic geovisualization techniques. In this paper, a new methodological framework for measuring and visualizing place-based job accessibility in space and time is presented that overcomes these three limitations. First, discretization and dasymetric mapping approaches are used to disaggregate counts of jobs and workers over specific time intervals to a fine-scale grid. Second, Shen's (1998) gravity-based accessibility measure is modified to account for temporal fluctuations in the spatial distributions of the supply of jobs and the demand of workers and is used to estimate hourly job accessibility at each cell. Third, a four-dimensional volumetric rendering approach is employed to integrate the hourly job access estimates into a space-time cube environment, which enables the users to interactively visualize the space-time job accessibility patterns. The integrated framework is demonstrated in the context of a case study of the Tampa Bay region of Florida. The findings demonstrate the value of the proposed methodology in job accessibility analysis and the policy-making process.

# Introduction

In past decades, much research effort has been devoted to studying accessibility, such as what it means, how to measure it, and what policy implications it offers. As a

---

concept discussed in a number of disciplines, such as transportation, urban planning, geography, and environmental policy, accessibility has a variety of definitions (Geurs and van Wee, 2004; Le Vine et al., 2013; Song, 1996; Shen, 1998). For example, it measures the potential of various opportunities for interaction (Hansen, 1959), the relative ease with which opportunities can be reached from a given location with a choice of travel (Morris et al., 1979), the potential of an individual to reach opportunities (Burns, 1979), and the benefits an individual gains from a given transportation or land-use system (Ben-Akiva and Lerman, 1979). Regardless of the definition, opportunities can be employment, transportation, education, food, health care, landscape and natural resources, or other facilities (e.g., Antrop, 2004; Apparicio et al., 2007; Ikram et al., 2015; Luo and Wang, 2003; Murray and Wu, 2003; Rigolon, 2016; Shen, 1998; Talen, 2001). Owing to the strong link to urban planning and economic development, an emphasis on *job accessibility* has long been recognized (Cervero, 1989; Ihlanfeldt, 1993; Kain, 1968; Matas et al., 2010; Wang and Minor, 2002).

Accessibility measures can be infrastructure-based, place-based, person-based, and utility-based (Gerus and Wee, 2004). In terms of job accessibility, a commonly-accepted notion refers to that it measures the relative ease with which job opportunities can be reached from a given location with a choice of transportation mode, and hence place-based measures are more appropriate. Place-based measures are aggregated metrics that assess the number of job opportunities that are accessible to workers from a given location or zone. These include jobs-housing balance ratios, cumulative-opportunity measures, and gravity-based measures. Though they do not capture individual components of accessibility like person-based measures (Kwan, 1998; Miller, 1991), place-based measures reflect accessibility at a zonal/regional level and hence are more favored by policy-makers (Dodson et al., 2007). Of the three groups of place-based accessibility metrics, gravity-based measures are generally most preferred, because they consider more components of accessibility, including the supply of jobs, the demand of workers, and the spatial barriers that separate them.

Hansen (1959) was the first effort that applied the gravity-based model to measure job accessibility. This model consists of two major components—number of job opportunities available in an employment zone and travel impedance between residential and employment zones. It is formulated as:

$$A_i = \sum_{j=1}^{n} S_j f(d_{ij}) \quad (1)$$

where $A_i$ is the job accessibility for zone $i$, $S_j$ is the number of jobs available in zone $j$ ($n$ represents the total number of employment zones), $d_{ij}$ is the travel impedance (distance or time) for a trip between $i$ and $j$, and $f(d_{ij})$ is a distance decay function (e.g., power function, exponential function, and Gaussian function) capturing how job accessibility changes relative to the distance. The integration of the distance decay effect enables gravity-based measures to consider all trip end possibilities and thus to avoid the so-called edge effect (Cervero et al., 1995). Despite being more meaningful than other approaches, Hansen's gravity-based measure lacks the consideration of the competition between demands, that is, workers, for job opportunities. The impact of demands on job accessibility is absent in Hansen's measure, indicating that a zone's accessibility to employment is irrelevant to the number of demands competing in it. When controlling the number of jobs available in zones and travel impedance between zones, job accessibility for a zone, apparently, will decrease if the number of workers living in the zone expands. Omitting impacts of demand can therefore lead to misleading results. This issue was later addressed by Shen (1998) who incorporated demand potential into Hansen's measure:

$$A_i = \sum_{j=1}^{n} \frac{S_j f(d_{ij})}{D_j} \quad (2)$$

$$D_j = \sum_{k=1}^{m} P_k f(d_{kj}) \quad (3)$$

where $D_j$ is the demand potential in zone $j$, which is captured by $P_k$, the number of workers living in zone $k$ ($m$ represents the total number of residential zones) and $d_{kj}$, the travel impedance between zone $k$ and zone $j$. All other symbols are the same as Equation 1. Since Shen's measure of accessibility accounts for the demand potential in each zone, it is found that this measure generally outperforms other ones (Song, 1996); for example, Merlin and Hu (2017) reported that the job accessibility measures derived from Shen's model were strongly related to employment rates than other models without considering demands.

This widely applied gravity-based measure is not free of concerns, however. First, it is usually applied in such a way that all three components of job accessibility—supply, demand, and spatial barrier—are time irrelevant. Nonetheless, they are subject to change in reality. For instance, the number of jobs that start in the morning is different from that in late evening. Likewise, most workers are not home all day long, and some may prefer leaving for work in early morning while others prefer afternoon or evening. Similarly, spatial barriers such as travel times often fluctuate temporally. Clearly, omitting such temporal dynamics in job accessibility could also result in inaccurate or misleading estimates. Though well addressed in person-based accessibility measures, temporal instability has received little attention in place-based job accessibility studies. There are a few attempts examining the *long-term* trend of job accessibility patterns. Ottensmann (1980) investigated the changes in job accessibility in the Milwaukee urban area in Wisconsin from 1927 to 1963 using a gravity-based accessibility measure. Using the same measure, Cervero et al. (1999) examined the shifts in job accessibility in the San Francisco Bay Area from 1980 to 1990. Instead of a gravity-based measure, Holloway (1996) used mean commuting time to measure job accessibility changes between 1980 and 1990 in large US metropolitan areas. These efforts mainly explored job accessibility changes over long periods of time but did not consider variations of the supply and demand over short temporal intervals. One exception considered daily fluctuations in the measurement of job accessibility by transit in the Greater Toronto Hamilton Area, Canada (Boisjoly and El-Geneidy, 2016), but it relied on a cumulative opportunity approach that did not incorporate demand. Consequently, attempts to integrate short-term instability in job accessibility measures are imminently needed. The U.S. Census Bureau's Census Transportation Planning Products (CTPP) report the number of workers and jobs at multiple aggregations (e.g., traffic analysis zone, TAZ) by commuters' time leaving homes and arriving jobs, respectively. This paper uses this data set to retrieve the hourly distributions of workers and jobs in TAZs.

The second concern of existing job accessibility measures is related to data aggregation and scale effect. Oftentimes, the above measures are applied to demographic and economic data aggregated at multiple geographic scales such as counties, census tracts, TAZ, and census blocks. Primarily designed for census purposes, these political or enumeration units may not appropriately capture job distribution patterns. In addition, according to the well-known modifiable analysis unit problem (Openshaw, 1984), job accessibility estimates by the place-based measures are likely to vary when the size and/or the configuration of the chosen analysis unit are changed. Altering analysis unit from one scale to another, e.g., from census tracts to blocks, triggers the change of zone size. Altering zone schemes while controlling the number of zones brings in the change of zone configuration. Furthermore, the spatial barrier (distance or time) between zones is typically estimated between zone centroids, which assumes that all workers and jobs in a zone are located at its center. This process may lead to significant aggregation errors, especially when the zone is large (Apparicio et al., 2008; Hewko et al., 2002; Hu and Wang, 2016). Using a geographic unit that is comparable (of similar size) can help mitigate the above impacts. In this research, a fine-scale grid is created as the analysis unit, and the dasymetric mapping is then applied to redistribute the hourly counts of workers and jobs associated with TAZs to the grid cells for more accurate and reliable estimates. These hourly estimates are then fed into a modification of Shen's model—the Generalized Two-Step Floating Catchment Area (G2SFCA) method—to measure hourly job accessibility at each grid cell.

The third concern refers to the lack of dynamic geovisualization methods for mapping place-based job accessibility for multiple time periods. Common approaches create separate choropleth accessibility maps for each time period and then either arrange them in rows or generate an animation for visualizing spatio-temporal variations of job accessibility. In either setting, spatial and temporal data are displayed in separate views. A number of accessibility maps are discretely visualized and compared, omitting the transition of job access between two times and potentially concealing the actual patterns (Goovaerts, 2010). Accordingly, this research applies a four-dimensional volumetric rendering approach to integrate the hourly job access estimates into a space-time cube environment, which enables users to interactively visualize space-time job accessibility patterns.

Though not related to the measurement of job accessibility, a few recent studies have attempted to develop time-varying accessibility indicators by considering dynamic demand, supply, or transportation system. For example, Järv et al. (2018) measured the time-varying food accessibility based on residents' travel times by public transit and walking to the closest food store in Tallinn, Estonia. In their model, mobile phone call detail records were used as a proxy for the dynamic demand; grocery store opening hours were considered to reflect the supply availability; and General Transit Feed Specification (GTFS) data were employed incorporate the travel time fluctuations. Lee et al. (2018) estimated the time-varying accessibility to bus services in Seoul, South Korea using a Huff model-based floating catchment area method. They used mobile phone call detail records and bus schedule data to approximate the dynamic demand and supply, respectively. But their work failed to incorporate the dynamic transportation system by only using travel distances. Using Hansen's gravity-based model (Equation 1), Moya-Gómez et al. (2018) measured the accessibility to the overall services in Madrid, Spain. Despite demand being absent, Twitter data and historical travel times from TomTom (https://www.tomtom.com) were employed in their model to account for the time-varying supply and traffic conditions, respectively. Built upon the space-time utility functions, Wang et al. (2018) examined the accessibility to food stores in Wuhan, China. Their model considered dynamic traffic conditions and supply attractiveness using taxi cab GPS data and stores' rating data, respectively. Despite these recent methodological efforts in developing dynamic accessibility, the three aforementioned concerns still have not been completely addressed in any of these studies. Specifically, these studies, except Lee et al. (2018), neglected some components of accessibility in their models. For instance, Järv et al. (2018) only considered the travel time, and Moya-Gómez et al. (2018) further incorporated the supply in their studies. In contrast, all three components of accessibility—supply, demand, and spatial barrier—can be considered in the proposed model. In addition, only a few temporal snapshots of accessibility maps were used to identify the spatio-temporal variations of accessibility in these studies, such as morning rush hour, evening rush hour, and late night in Lee et al. (2018), morning peak, midday, and afternoon peak in Moya-Gómez et al. (2018), and breakfast, lunch, and dinner time in Wang et al. (2018). In this research, a space-time cube environment is developed for more fully understanding spatio-temporal patterns of accessibility. Furthermore, they all utilized big data to capture the temporal dynamics in the components of accessibility. Specifically, mobile phone call detail records were used as a proxy for the demand in Järv et al. (2018) and Lee et al. (2018), Twitter data as the surrogate for service attractiveness in Moya-Gómez et al. (2018), and TomTom data and taxi GPS trajectory for traffic conditions in Moya-Gómez et al. (2018) and Wang et al. (2018), respectively. While big data like mobile phone call records have finer spatial and temporal granularity, they are not as freely available to the public as the census data. Finally, the present study focuses on measuring accessibility to employment, whereas the above studies on other service types such as public transit and foods.

The aim of this research is to propose a methodological framework for measuring and mapping place-based space-time job accessibility. It addresses the above three issues that are crucial to the study of job accessibility (and beyond). Despite that each of the applied methods (e.g., dasymetric mapping, G2SFCA, and volumetric rendering) has been separately developed and discussed in previous studies, to the best of our knowledge, no research has combined these methods into one framework to measure and map place-based space-time job accessibility. The integrated framework is demonstrated in the context of a case study of the Tampa Bay region of Florida. The methods can help inform transportation planning (e.g., transit management), economic development (e.g., affordable housing), and healthcare (e.g., EMS) resource planning and adjustment.

**Methods**

A workflow describing the methods for computing the place-based space-time job accessibility measure is illustrated in Figure 1. This framework is best described according to the chronological order of the workflow: (1) definition of study area and input data, (2) discretization of time and space, (3) calculation of space-time accessibility that is adapted from Shen's (1998) measure and implemented using a G2SFCA process, and (4) geovisualization of the results. Each of these steps are described in detail below.

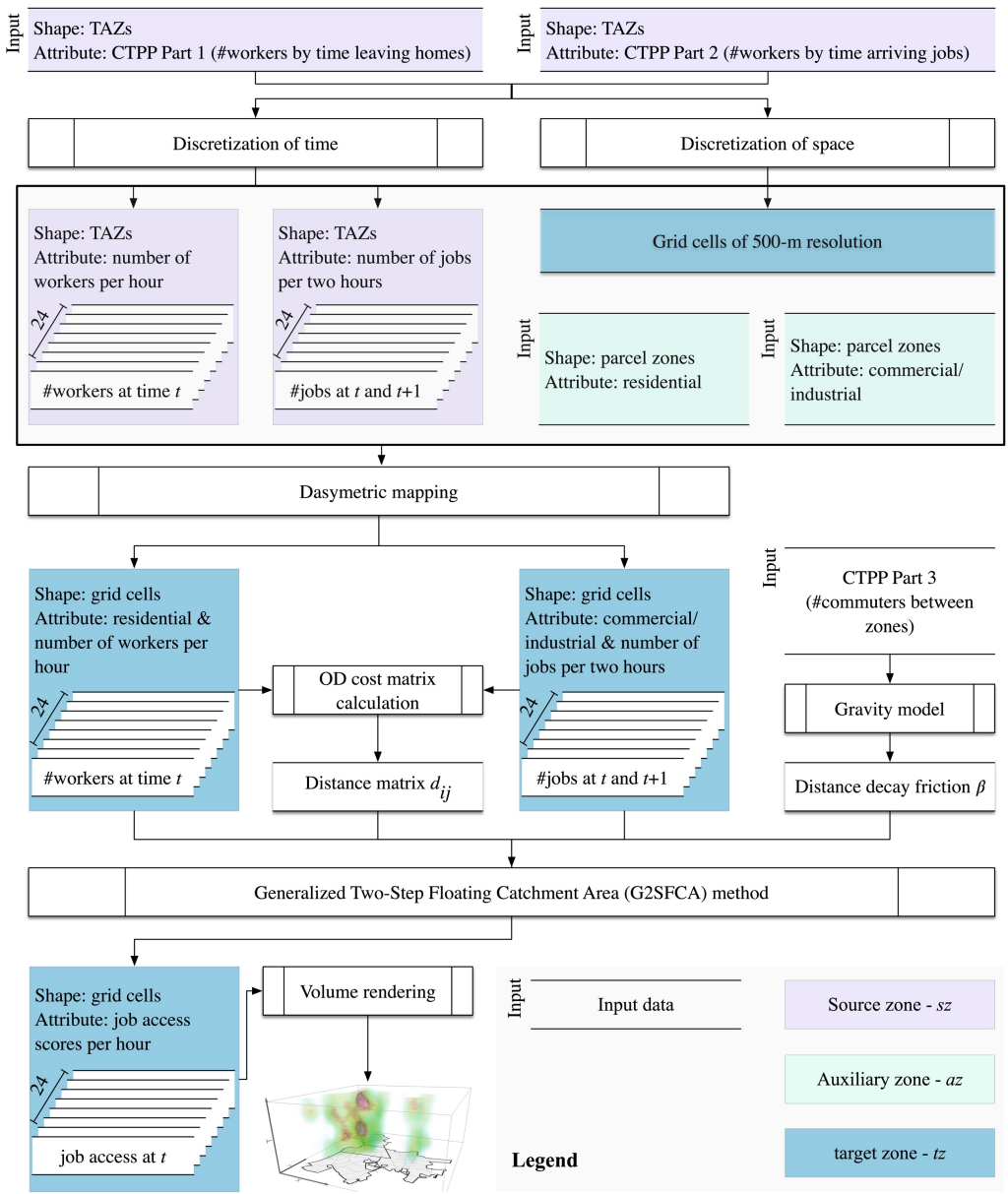

Figure 1. Workflow of the proposed methodology

*Study Area Definition and Data Sources*
Hillsborough and Pinellas Counties in the Tampa Bay Area of Florida were selected as the study area. Several major cities including Tampa, St. Petersburg, and Clearwater are within this area. Along with Hernando and Pasco Counties to the north, they make up the Tampa–St. Petersburg–Clearwater Metropolitan Statistical Area. According to the 2006-2010 American Community Survey (ACS) 5-Year Estimates data, there were 564,303 workers living in Hillsborough County; 89.5 percent of them stayed in that county for work and 5.2 percent commuted to Pinellas County. For Pinellas County, of its 414,241 resident workers in total, 87.1 percent stayed for work and 1.1 percent commuted to Hillsborough County.

The major data source used in this research is the most recent 2006 to 2010 CTPP data based on the 5-year ACS (http://ctpp.transportation.org/Pages/5-Year-Data.aspx). The CTPP consists of three parts: Part 1 reports the number of workers at residential places (total number, breakdowns by socioeconomic variables such as age, wage, and time leaving home); Part 2 records the number of workers at workplaces (total number, breakdowns by socioeconomic attributes such as wage and time arriving jobs); Part 3 provides detailed journey-to-work flow (total number of commuters, mean travel time by specific transportation modes) between residential and employment places (Hu and Wang, 2015). Temporal frequency distributions of workers and jobs, critical components to the measurement of space-time job accessibility, are retrieved from the number of workers by their times leaving homes (Part 1) and number of workers by their times arriving jobs (Part 2), respectively. They are provided in fifteen-minute increments from 5:00 a.m. to 11:00 a.m., one-hour increments from 11:00 a.m. to midnight, and a five-hour increment from midnight to 5:00 a.m. The data are aggregated at multiple zone levels such as Metropolitan Statistical Area (MSA), Census Tract, Traffic Analysis District (TAD), and TAZ. This research uses the TAZ-level data since TAZs are the most disaggregate zone unit in the CTPP, and they are considered to be relatively uniform in terms of land use. There are 1,283 TAZs (compared to 567 census tracts) in the study area in 2010, averaging 3.78 km$^2$ in size. The TAZ boundary and other spatial datasets such as the road network were extracted from the TIGER Products 2010 from the U.S. Census Bureau. See Figure 2 for a map of the study area, where the major job centers, major roads, census tract boundaries, and the commuting flow patterns between tracts are shown. It should be noted that the use of tracts instead of TAZs in the visualizations is only for the sake of better visualization, since it would be too crowded to observe any patterns if mapped at the TAZ level.

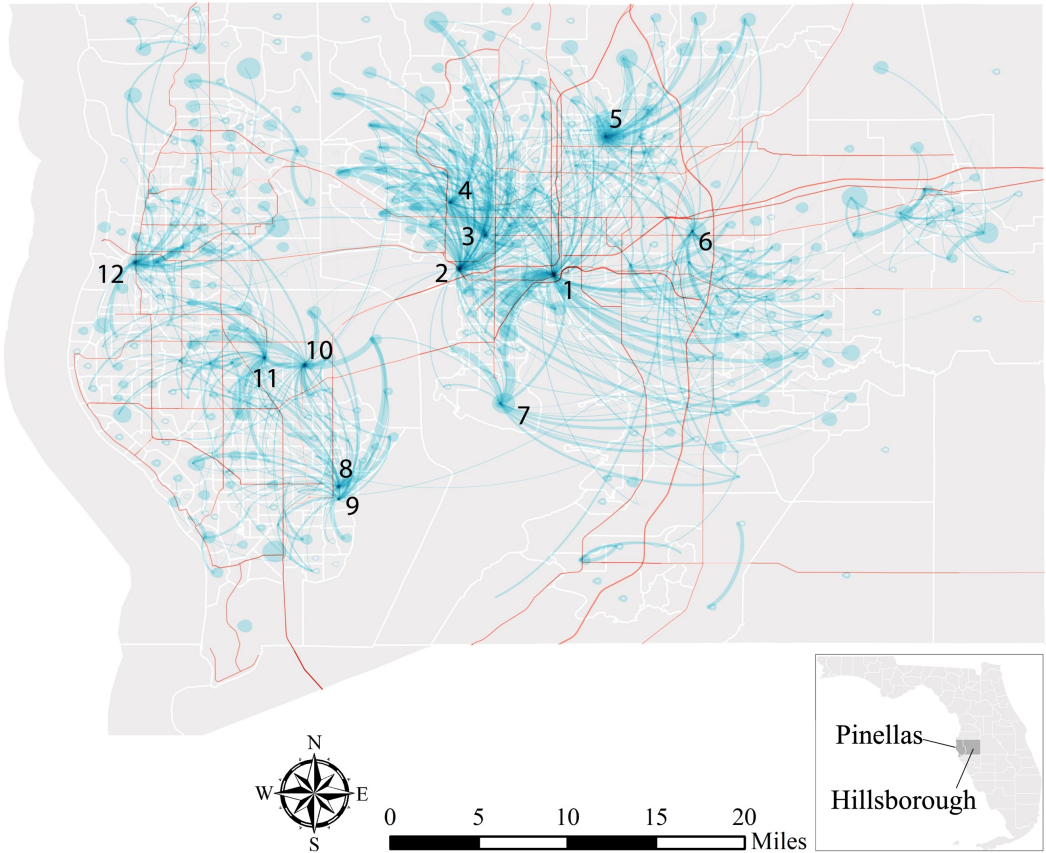

Figure 2. Census tracts overlaid with major roads and commuting flow in the study area. Major job centers are highlighted and labeled: 1. Tampa CBD, 2. Westshore Plaza, 3. International Plaza and Bay Street Shopping Mall, 4. USCIS Tampa Office, 5. University of South Florida, 6. Campuses of Hillsborough Community College, Springfield College, and Jersey College Nursing School, 7. MacDill Air Force Base, 8. St. Petersburg CBD, 9. University of South Florida St. Petersburg Campus, 10. Jabil Company (a US-based global manufacturing services company), 11. Pinellas Park (a major Gateway area of Pinellas County with many marine businesses), 12. Clearwater City Hall. [Note: The larger the line the more commuters, and lines with fewer than 100 commuters were cut off for better visualization.]

*Discretization of time and space*

The first step of the proposed methodology is to generate counts of jobs and workers at consistent, fine-scale spatial and temporal resolutions. For this study, this involved discretizing both the CTPP Part 1 (number of workers by time leaving homes) and Part 2 data (number of workers by time arriving jobs) into consistent representations in both time (hourly interval) and space (grid cells). In terms of time, CTTP counts of jobs and workers were either aggregated or disaggregated to hourly measures following Kobayashi et al. (2011)'s study. For data from 5:00 a.m. to 11:00 a.m., corresponding 15-min counts were summed to obtain hourly estimates. For data from 12:00 a.m. to 5:00 am, the five-hour count was evenly divided into 1-hour intervals. For 11:00 am to 12:00 a.m., the raw hourly counts were used. For example, if fifty people in TAZ *A* left homes for work every fifteen minutes from 6:00 a.m. to 7:00 a.m., there would be a total of 200 resident workers in TAZ *A* for the whole hour. If 150 of those commuters arrived at work in TAZ *B* between midnight and 5:00 a.m., there would be thirty jobs assigned to each of the five hours in TAZ *B*. In this way, 24 groups of hourly counts of workers $wt_i$ and jobs $jt_i$ ($i = 1,..., 24$) were obtained for each TAZ. In terms of space, a regular grid was imposed over the study area for the purposes of redistributing the hourly count data associated with each TAZ to a uniform, finer scale in order to overcome issues associated with the large sizes of the TAZs and the potential uneven distribution of jobs and workers within them. For the study area, a raster of 22,646 grid cells of 500-m resolution was tessellated over the study area. This configuration was selected by trial and error in an attempt to balance accuracy and computational efficiency (i.e. a larger number of cells improves accuracy but demands more computation power).

    The second step uses dasymetric mapping to redistribute the hourly count data associated with each TAZ to the grid cells. Dasymetric mapping is an areal interpolation method that utilizes auxiliary data such as land use to disaggregate population data from a coarse spatial resolution to a finer resolution (Mennis and Hultgren, 2006). Owing to its integration of auxiliary data in the process, dasymetric mapping can provide more accurate small-area population estimates than other techniques not using such auxiliary data (Gregory, 2002). Parcel-based land use data of the study area in 2010 provided by the Florida Department of Revenue were used for this purpose. The data set included several land use categories, such as residential, commercial, industrial, agricultural, and water. In total, there were 49,269 residential parcels and 18,703 commercial/industrial parcels in the study area. The former was used to calibrate where workers live and the latter to adjust jobs' spatial arrangement. Specifically, dasymetric mapping was applied to estimate the numbers of workers and jobs in each grid cell for each time *t*. Following the notions in Mennis and Hultgren (2006), the TAZs are referred to as *source zones* (*sz*), land use polygons are termed *auxiliary zones* (*az*), and the grid cells are called *target zones* (*tz*). The goal is to derive the numbers of workers and jobs at each *tz* based on the spatial overlap of *sz*, *az*, and *tz*. Equation 4 formulates the above process:

$$\hat{y}_{tz} = \sum_{sz \in tz} \left( \frac{y_{sz}}{\sum_{az \in sz} A_{az}} \times \sum_{az \in sz} A_{az \cap tz} \right) \quad (4)$$

where  $\hat{y}_{tz}$ = the estimated count of workers (or jobs) for the target zone *tz*,

$y_{sz}$ = the recorded count of workers (or jobs) of a source zone $sz$ that overlaps $tz$,

$A_{az}$ = the area of a given residential land use zone $az$ (or a commercial/industrial land use zone to estimate jobs),

$A_{az \cap tz}$ = the area of the intersection between $az$ and $tz$.

Figure 3 illustrates the process of dasymetric mapping using an example of the reported number of workers who leave home for work between 7:00 and 8:00 a.m. in TAZ 00136022 in the study area. Within the parenthesis in Equation 4, the first component of the product calculates the density of workers in a $sz$ (a TAZ depicted in purple); instead of the whole area of a $sz$, the areas of all residential zones $az$ (represented in green) within a $sz$ are used to eliminate uninhabitable areas. Therefore, the density represents the number of workers per square kilometer of residential parcels in a TAZ. The second component computes the area of all residential parcels (or portions) associated with a $sz$ in a $tz$ (a cell outlined in dashed lines). Hence, the product returns the number of workers related to a $sz$ in a $tz$. As a $tz$ may contain regions belonging to multiple $sz$s, estimated workers for all $sz$s in a $tz$ are summed. In this way, the counts of workers are redistributed to a set of grid cells from TAZs. Likewise, the counts of jobs at each grid cell can be obtained. After the dasymetric mapping process was applied to the study area, cells located outside residential and commercial/industrial areas were excluded, with 9,965 remaining residential cells and 5,898 employment cells (in comparison to 1,283 TAZs for either workers and jobs).

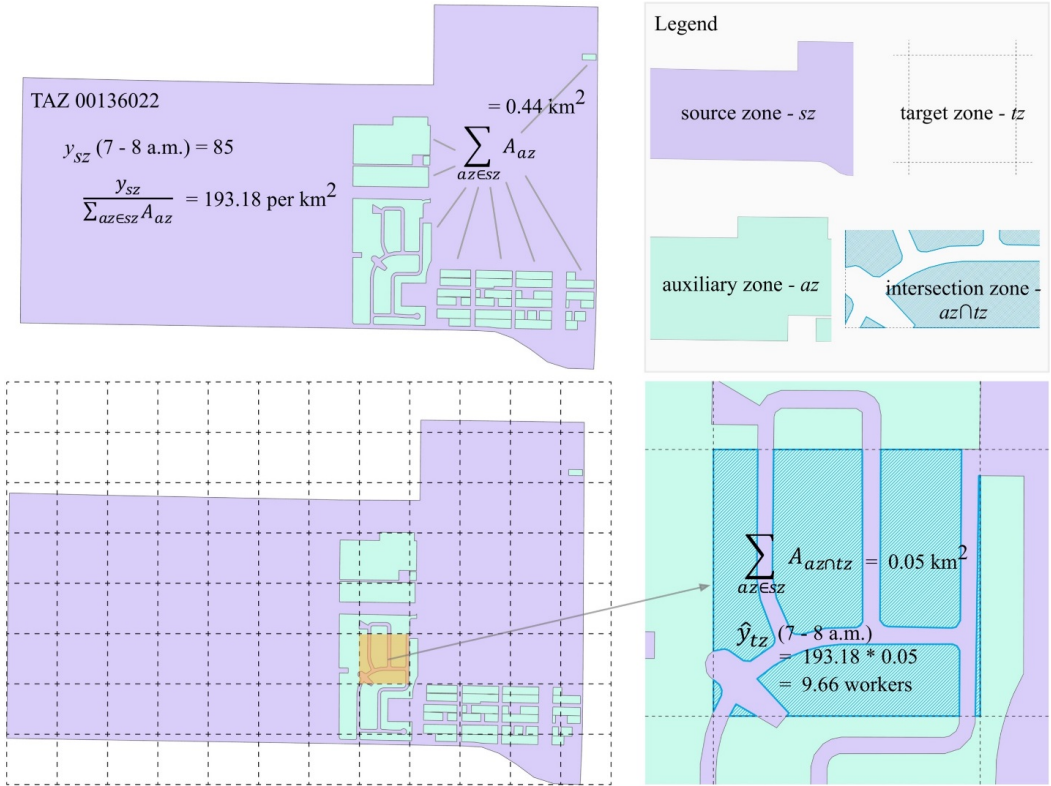

Figure 3. Illustration of the dasymetric mapping process

*Measurement of space-time job accessibility*

The third major step in the workflow involves modifying Shen's (1998) accessibility measure to incorporate a temporal dimension into the gravity equations. Shen's (1998) measure was adapted using the G2SFCA process (Wang, 2012; 2015) which simplifies the calculation by dividing it into two steps. In the context of job accessibility, the first step simply calculates the supply-to-demand ratio around each job location, while the second step sums up the ratios associated with all supply locations that are accessible to a demand location. Hence, G2SFCA can be interpreted as the ratio between supply of jobs and demand of workers and is equivalent to Shen's measure. Due to its simplified structure and interpretation, G2SFCA is more often used in spatial accessibility studies, in particular, access to health care research (e.g., Guagliardo, 2004; Ikram et al., 2015; Luo and Wang, 2003).

Based on the hourly counts of workers and jobs as well as the travel costs between them, the G2SFCA model was adapted to compute job accessibility scores for each time $t$. The first step of the revised model simply calculates the supply-to-demand ratio $R_j$ around each job location $j$ at time $t$ using the following formula:

$$R_{j,t} = S_{j,t\in[t,t+1]} \Big/ \left(\sum_{k=1}^{m} D_{k,t} f_t(d_{kj,t})\right) \quad (5)$$

where $D_{k,t}$ represents the number of resident workers at location $k$ at time $t$, $m$ is the total number of residential locations, $S_{j,t\in[t,t+1]}$ denotes the number of jobs at location $j$ at time $t$ and the following time $t+1$ ($t+1$ was added to allow for longer travel times across the study area); for example, resident workers who leave home between 7:00 and 8:00 a.m. can access jobs with a start time range from 7:00 to 9:00 a.m., $d_{kj,t}$ is the travel time between location $k$ and $j$ at time $t$, and $f_t(d)$ defines a continuous distance decay function such as power, exponential, and Gaussian functions for time $t$. Since a particular residential location could have access to multiple job locations, a second step is then followed to sum up the ratio $R_{j,t}$ at any job location $j$ that could be reached from a residential location $i$ at the same time $t$:

$$A_{i,t} = \sum_{j=1}^{n} R_{j,t} f_t(d_{ij,t}) \quad (6)$$

where $n$ represents the total number job locations and $A_{i,t}$ denotes the job accessibility at residential location $i$ at time $t$. The same $f_t(d)$ function is also integrated to take into account the continuous decay of access in distance between $i$ and $j$ at time $t$. Plugging $R_{j,t}$ into Equation 6, $A_{i,t}$ is then transformed as

$$A_{i,t} = \sum_{j=1}^{n} \left[ S_{j,t\in[t,t+1]} f_t(d_{ij,t}) \Big/ \left(\sum_{k=1}^{m} D_{k,t} f_t(d_{kj,t})\right) \right] \quad (7)$$

and a larger value of $A_{i,t}$ indicates a better job accessibility at residential location $i$ at time $t$.

Here, the centroids of $tz$s (again, grid cells) are used to represent location $i$ (or $k$) and $j$, network distances between cell centroids are employed to define the travel cost $d_{ij}$, and a power function is utilized to account for the distance decay effect as it is more suitable for analyzing short distance interaction at the city scale than other decay functions (Fotheringham and O'Kelly, 1989, pp. 12-13). The decay friction coefficient $\beta$ is estimated using a gravity model (Hu and Wang, 2016; Wang and Minor, 2002; Wang, 2003):

$$C_{ij} = D_i S_j d_{ij}^{-\beta} \quad (8)$$

where $C_{ij}$ represents the number of commuters residing in zone $i$ and working in zone $j$, $D_i$ and $S_j$ are the number of workers in zone $i$ and number of jobs in zone $j$, respectively. $C_{ij}$ are extracted from CTPP Part 3, $D_i$ and $S_j$ are obtained from CTPP Part 1 and Part 2, respectively. For the study area, 24 groups of hourly counts of workers (e.g., 12:00 – 1:00 a.m., 1:00 – 2:00 a.m., ..., 11:00 p.m. – 12:00 a.m.) and jobs (e.g., 12:00 – 2:00 a.m., 1:00 – 3:00 a.m., ..., 11:00 p.m. – 1:00 a.m.) for each grid cell $tz$ were fed into the G2SFCA model for measuring the hourly job accessibility across Tampa Bay. This leads to an o-d cost matrix for 9,965 * 5,898 = 58,773,570 pairs to be solved and fed into the G2SFCA model for each time $t$. Since it would be unrealistic to obtain travel times based on real-time or historical traffic data for such a massive matrix, instead, network distances for all o-d pairs were calculated, and a distance decay friction coefficient $\beta$ was estimated to be -0.602 by using Equation 8 in order to calculate the accessibility measures. See the subsequent section for an exploratory analysis that adopts Google Maps travel times for job accessibility measurement.

*Visualization of space-time job accessibility*
The final step in the workflow involves creating a dynamic geovisualization of the hourly accessibility scores from the previous tasks. The resulting space-time job accessibility has a four-dimensional volume structure, $(x, y, t, A_{x,y,t})$, where $A_{x,y,t}$ denotes the job accessibility for location $(x, y)$ at time $t$. Some popular approaches to the visualization of such four-dimensional data include: (1) volume rendering that assigns color and transparency to space-time cells (or, voxels) based on their accessibility scores (cells with poorer job access are less visible while high job access cells are more observable with different colors) and (2) isosurfacing that connects together points with the same accessibility value into a volume (Brunsdon et al. 2007; Demšar and Virrantaus 2010; Hu et al., 2018). The techniques have been commonly used to visually analyze spatio-temporal patterns such as human mobility (Kwan, 2000) and crime (Hu et al., 2018; Nakaya, 2013). Compared to the traditional side-by-side visualization of a series of two-dimensional maps, a four-dimensional visualization setting might allow for a more effective and interactive way to visualize and detect space-time dynamics of job accessibility. Here, a volumetric rendering technique is applied to combine the job accessibility estimates across different times through a trilinear interpolation and plot the results in a space-time cube environment. In this interactive environment, users can easily query data, alter the attributes of a scene (e.g., viewing angle and voxel transparency), add elements (e.g., contour lines), and see the results of these actions in real time. However, the two-dimensional displays of the space-time cube environment in this paper would inevitably conceal some patterns that can be observed in the space-time environment (Kwan, 2000). For the study area, space-time cubes were used to map hourly job accessibility. The space-time cubes consisted of (some number) of voxels.

For comparison to the space-time cube representations, job accessibility was also measured using the original G2SFCA (static) implementation of Shen's (1998) model (refer to Wang, 2012 for the equations). Specifically, both the hourly counts of workers and jobs from the dasymetric grid were summed to derive daily counts at each grid cell. The daily counts, distance decay coefficient obtained from Equation 8, and the o-d distance matrix derived above were then fed into the original G2SFCA model to calculate the static job accessibility.

**Results and Discussion**

When applying Shen's (1998) traditional gravity-based measure, job accessibility within the Tampa Bay region exhibits a concentric distribution pattern, decreasing gradually from the peninsular towards inland areas (Figure 4). Notably, the Tampa peninsular area has the best job accessibility due to the high-density of jobs and convenient, highly-developed transportation networks. Also evident from the map is a spillover pattern of high job access into areas that are immediately accessible from the Tampa peninsular through major highways. The extent of high job access areas largely matches the high-density employment areas identified in Figure 2. Inland areas, instead, have more residential than commercial/industrial activities and thus have lower job accessibility. This pattern is also present in Figure 2—a large number of commuting trips to the job centers start from inland suburban areas.

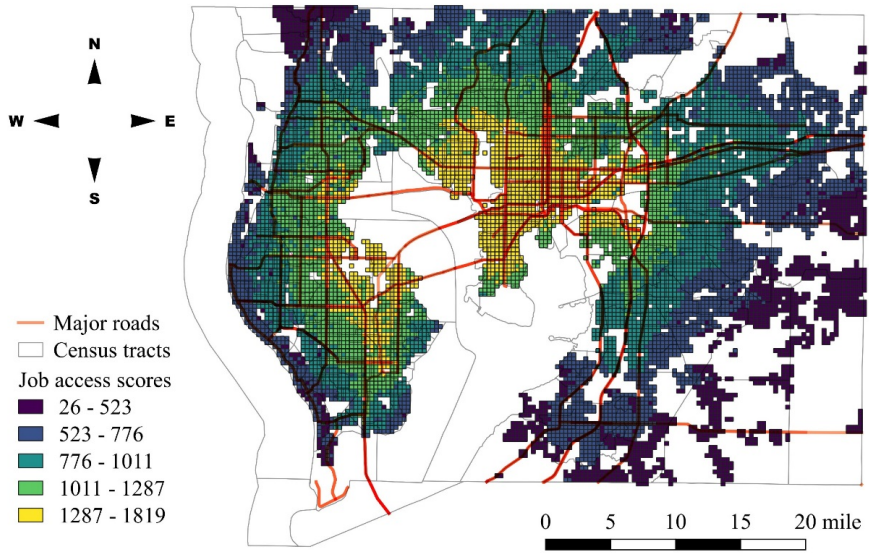

Figure 4. Job accessibility pattern measured by the static model

Although the traditional method illustrates broad spatial patterns of job accessibility in the region, the space-time approach presented here reveals the dynamic changes over the course of a day (Figure 5). Specifically, Figure 5a displays the pattern using the volumetric rendering approach that adjusts the color and transparency of each voxel based on its job accessibility score. As expected, job accessibility not only varies from place to place but also shifts throughout the day. In general, job accessibility for the entire study area peaks at 5:00 – 7:00 a.m. (see the substantial areas of voxels with high job access in Figure 5a). As stated, it is assumed that a worker can access jobs in the current two-hour window. Therefore, workers who usually leave home around 5:00 – 7:00 a.m. have access to jobs starting between 5:00 and 8:00 a.m. In essence, the G2SFCA approach measures the ratio between jobs and workers. Despite that a large number of daytime jobs start between 7:00 and 9:00 a.m., the amount of competing demands during that time is the greatest as well. In comparison, the number of workers who leave home between 5:00 and 7:00 a.m. is much smaller relative to that of accessible jobs (starting between 5:00 and 8:00 a.m.). This big contrast gives rise to job accessibility reaching its highest point between 5:00 and 7:00 a.m. After this period, job accessibility in the whole region decreases significantly and remains at low levels in spite of two moderate spikes at noon (2:00 – 3:00 p.m.) and midnight (11:00 p.m. – 12:00 a.m.). The former period reflects job shifts starting from 4:00 p.m. (again, job access between 2:00 and 3:00 p.m. considers jobs starting between 2:00 and 4:00 p.m.) and the latter corresponds to night shifts starting at midnight.

Compared to the traditional two-dimensional maps, the space-time cube setting provides an interactive visualization environment, allowing for users to rotate the cube and examine results at any viewing direction. Figure 5b shows the top front view of space-time job accessibility in the region. The dark purple belt represents areas of the highest job accessibility (5:00 – 7:00 a.m.) and its spatial extent is largely consistent with the pattern spotted in Figure 4. Figure 5c illustrates the isovolumes generated using a threshold value of the 95th percentile of the space-time job accessibility scores. Consistent with Figure 5a, the same three major periods of high job accessibility in the entire region are identified. The substantial shrinkage of the spatial extent from the morning commuting peak to the noon and further to the midnight job shift hours is especially noticeable. None of these dynamic patterns can be readily diagnosed by the conventional static job accessibility models.

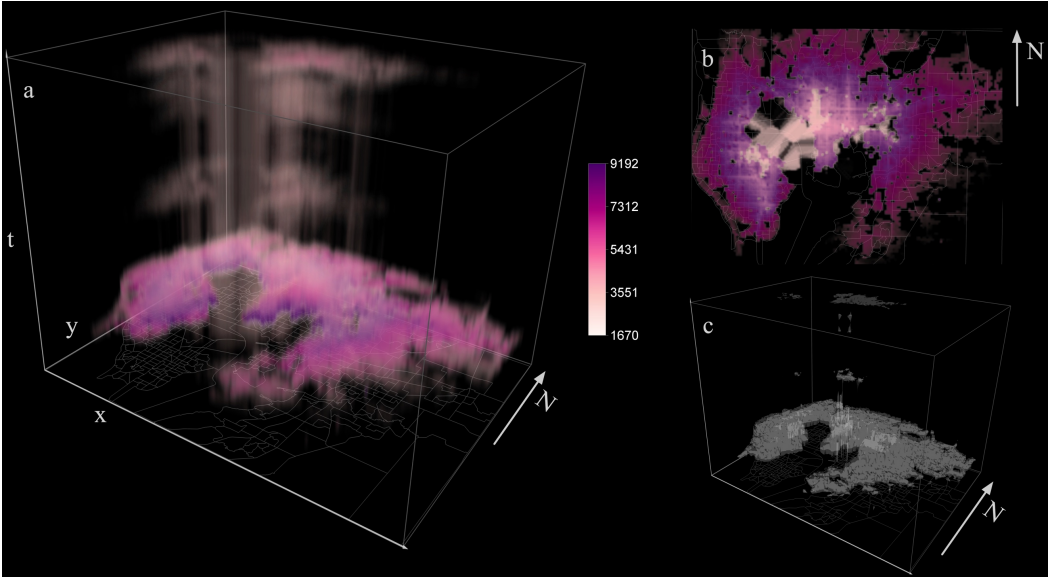

Figure 5. Space-time job accessibility pattern: (a) 135° viewing direction, (b) top front view, and (c) isovolumes generated using a job accessibility value of the 95th percentile

To understand to how much extent time could affect the measurement of job accessibility, four different modeling scenarios are highlighted: (1) static jobs and workers (Figure 4), (2) dynamic jobs and static workers (Figure 6a), (3) dynamic workers and static jobs (Figure 6b), and (4) dynamic jobs and workers in 24 hours (Figure 5a). Some interesting patterns are spotted and merit discussions. First, note the substantial differences in the magnitude of job accessibility scores of the four scenarios, especially between scenarios 2 and 3. As reported in Table 1, the mean job accessibility score of the entire region gradually increases from 63 in scenario 2 to 889 in scenario 1 and further to 2,573 in scenario 4, and then drastically rises towards 163,527 in scenario 3. Since scenario 2 employs dynamic jobs and static workers in the model, it is actually comparing the number of *hourly* job opportunities to that of *daily* workers. Understandably, this would lead to the lowest supply-demand ratio. Scenario 1, instead, switches job opportunities from hourly to daily estimates and experiences a 1,300% growth rate in the mean job accessibility. Scenario 4 considers the dynamics in both jobs and workers and roughly triples the mean job accessibility relative to scenario 1. It should be noted that significantly high job-to-worker ratios are found in some job centers shown in Figure 2 and overall, they drag up the mean value. This indicates that the conventional static model may underestimate job access as it tends to smooth out extremely high values. Scenario 3, exactly the inverse case to scenario 2, compares *daily* job counts to *hourly* workers and hence returns the highest mean job access score, which increases by 2,596 times relative to scenario 2.

Looking at Figures 6a and 6b allows us to visualize and further understand the differences between scenarios 2 and 3. For scenario 2, most of the region, particularly those job centers, has high job accessibility between 6:00 and 9:00 a.m. Given that this is the time in a day when the greatest number of job opportunities is available, especially for areas of high-density employment, this pattern makes sense. For scenario 3, two major temporal clusters of high job accessibility—one between 12:00 and 4:00 a.m. and another between 7:00 and 11:00 p.m.—are identified. As the supply side is fixed in scenario 3, the less the working population the greater the job accessibility. This explains why the two time periods stand out. In addition, areas of the highest job accessibility mostly appear in downtown area in the early morning cluster while the areas further spread out in the evening cluster. This perhaps indicates that the number of population working at night is the lowest in a day.

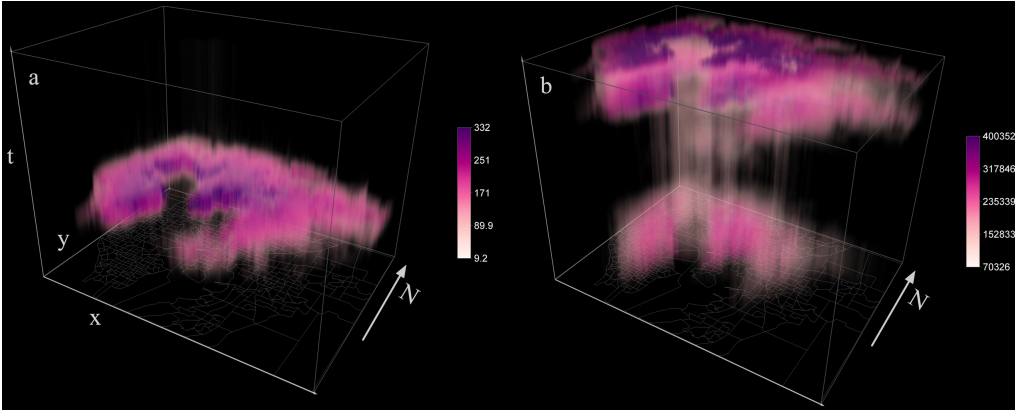

Figure 6. Space-time job accessibility pattern where (a) only jobs are dynamic and (b) only workers are dynamic

Table 1. Mean values and correlation coefficients of four modeling scenarios

|  | Mean values | Correlation coefficients | | | |
| --- | --- | --- | --- | --- | --- |
|  |  | Scenario 1 | Scenario 2 | Scenario 3 | Scenario 4 |
| Scenario 1 | 889 | - | - | - | - |
| Scenario 2 | 63 | 0.19*** | - | - | - |
| Scenario 3 | 163,527 | 0.40*** | -0.43*** | - | - |
| Scenario 4 | 2,573 | 0.50*** | 0.01*** | 0.33*** | - |

Note: *** indicates significant at the 0.001 level. Scenario 1: static jobs and workers; Scenario 2: dynamic jobs and static workers; Scenario 3: dynamic workers and static jobs; Scenario 4: dynamic jobs and workers.

A correlation analysis is followed to examine the statistical relationships between every two scenarios, and the results are reported in Table 1. Most interestingly, a relatively low positive correlation (0.5) between scenarios 1 and 4 was found. This indicates that the conventional static job accessibility model can only capture about 25% of the variance in space-time job accessibility, highlighting the need to incorporate time in measuring job accessibility. A significantly weak correlation (0.01) is found between scenarios 2 and 4, suggesting perhaps that a model where only jobs are dynamic could seldom improve the model accuracy. Instead, a model with only dynamic workers can somewhat improve the estimates (with a low correlation coefficient of 0.33). The negative sign between scenarios 2 and 3 reflects the expectedly inverse relationship between a job-dynamic model and a worker-dynamic model. It should be noted that the above correlation findings may only apply to the current case study. Further discussion about the differences among the four scenarios is beyond the scope of this paper.

The case study in Figure 5 captured the time dimension of employee and job availability but not that of the transportation system due to the complexity of gathering a massive o-d time matrix with traffic. An exploratory analysis using Google Maps Distance Matrix API was conducted to explore the effects of using more accurate network travel times on the model results. Specifically, the traffic-integrated travel time for each hourly group, say, 7:00 – 8:00 a.m. was measured by specifying the departure time (e.g., 7:00 a.m.) in the Distance Matrix API. Given the Distance Matrix API free usage limit (up to 40,000 o-d records per month, https://cloud.google.com/maps-platform/pricing/sheet/), the top ten tracts (156 residential cells) in terms of the number of workers commuting to downtown Tampa (8 employment cells) were selected and examined (see Figure 7e). In total, this selection leads to an o-d time matrix of 29,952 (156 * 8 * 24) records. These hourly travel times were then fed into the proposed model to measure the space-time job accessibility. Figure 7a shows the mean job accessibility measured by considering dynamic workers, jobs, and travel times for each of the ten tracts during the day, while Figure 7b illustrates the mean job accessibility where static travel distances were used. A largely consistent pattern was observed between Figures 7a and 7b, indicating that the use of travel distances was a reasonable choice in this study. As shown in Figure 7c, most of the ten tracts had relatively stable mean travel times throughout the day, resulting in a fairly stable, high correlation coefficient between dynamic times and static distances (except the morning peak time as shown in Figure 7d). The high correlation contributes to a similar job accessibility pattern in Figures 7a and 7b. However, compared to dynamic times, the use of static distances appears to slightly overestimate job accessibility as it omits traffic in the calculation.

This would be problematic in areas where traffic changes substantially over the day.

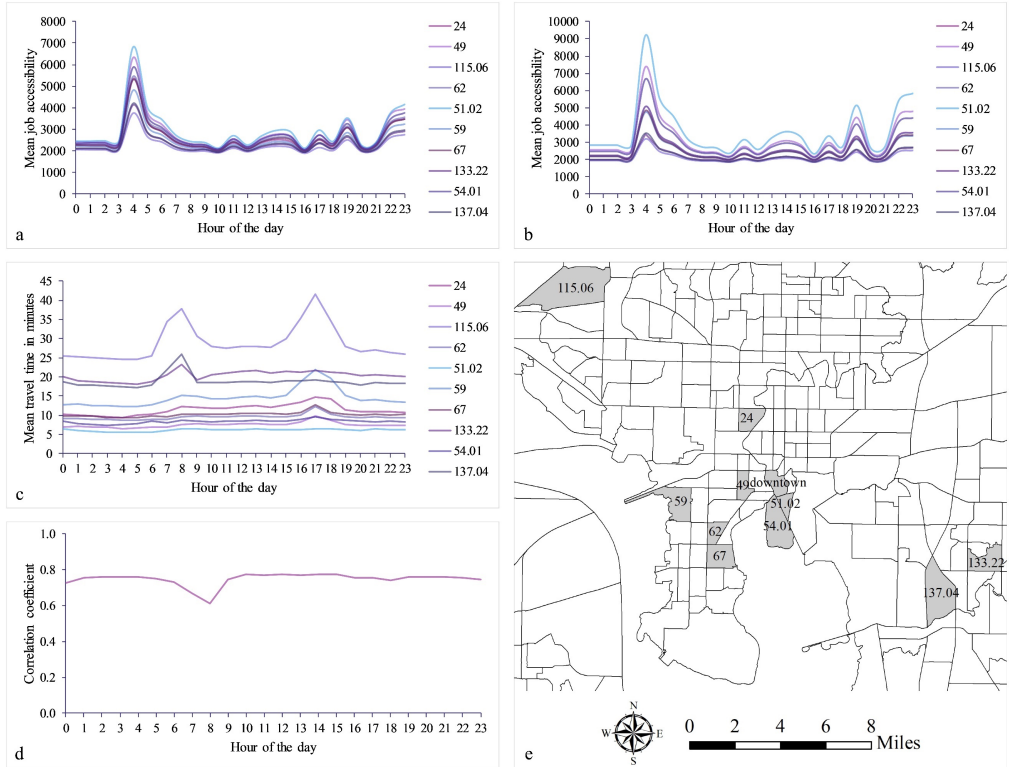

Figure 7. Mean job accessibility of ten selected tracts by the hour of the day, where (a) dynamic times and (b) static distances were used; (c) mean travel time of ten selected tracts to downtown Tampa by the hour of the day; (d) correlation coefficients between dynamic times and static distances by the hour of the day; and (e) locations of ten selected tracts.

## Conclusions

Job accessibility, an important metric to aid in decision-making in urban planning and economic development, is usually modeled by place-based measures such as the popular gravity model. Despite their prevalence usage, these conventional models are yet not free of concerns and may lead to ineffective policymaking. They assume stabilized supply and demand over time, rely on data often aggregated at large geographic scales, and identify temporal patterns by comparing separate maps of different time periods. This paper presented a new framework for measuring space-time job accessibility that overcame all of these concerns.

Although this paper specifically focuses on methodological contributions, it can bring useful policy implications to decision-makers. A central policy-related question this research aims to answer—but most conventional models fail to—is that whether high job access neighborhoods possess such privileges (or low job access communities keep suffering from such obstacles) throughout the day. In essence, the G2SFCA-derived job accessibility captures the (im)balance of jobs and workers, and the observed temporal variability of such an imbalance could help inform transportation planning, such as transit management and ride-sharing arrangement, and examine the impact of relevant policies including flexible work hours, mixed land use, and so on. The findings indicate that there is a significant variation in the temporal distribution of job accessibility in addition to the spatial variation in the study area. Three time periods of high job access in the Tampa Bay region are identified. Urban planners and policymakers should understand such variations in both space and time. For instance, planners can use the methods to identify any areas with persistent low job access. Policies aimed to bring more jobs appropriate for these residents close to them or improve their mobility by providing services like public transit or bike share may be beneficial. Likewise, researchers and planners can use the proposed methods to identify neighborhoods where job access fluctuated from high to low. It should be noted that the methods are not limited only to the examination of daily job accessibility. Other time-sensitive applications such as the examination of access to transportation, health care, and food can certainly benefit from this framework.

There are, however, some limitations to this study. As suggested by the exploratory analysis in Figure 7, it might be worthwhile to use dynamic travel times for the travel cost matrix since they can better capture the time dimension of the transportation system. Due to the complexity of attaining a massive real-time o-d time matrix in this study, the temporal variations of travel times were only examined in ten selected tracts. The proposed methods are capable of integrating all three dynamic components of job accessibility if these data are available. For example, studies of smaller scales may use Google Maps API for dynamic travel times and then feed into the proposed model. Studies of larger regions may use a larger cell size or the census zones to reduce the size of o-d cost matrix to a reasonable level that Google Maps API supports. Second, this research examined the access to all jobs for the general workers without accounting for the diversity of jobs and workers. Further studies may employ other data sources to classify jobs and workers into groups based on occupation or wage and then measure the group-specific space-time job accessibility patterns. Third, only automobile travel is examined in this study given it being the dominant mode of commuting in the study area (e.g., 94.3% of trips by automobile vs. 1.8% by public transit); however, it would be more meaningful to consider more transportation modes such as public transit in other study areas. Fourth, this study employed the dasymetric mapping technique to redistribute hourly count data in terms of jobs and workers in TAZs to grid cells for a sharper spatial resolution. Other areal interpolation approaches, such as the road network hierarchical weighting interpolation that assumes the population distribution being closely related to the street network (Xie, 1995) and the spatial statistics-based interpolation that accounts for spatial autocorrelation of population distribution (Murakami and Tsutsumi, 2011) could be tested. Finally, new sources of data such as cell phone activity or positioning data will be explored to improve this study. As the recorded locations in the data are the actual places instead of zones, such big data can help detect site-specific accessibility patterns. The high temporal resolution can benefit studies interested in detecting temporal variations of accessibility at shorter intervals. Furthermore, traffic congestion level of streets can be derived from the travel speeds associated with each waypoint in the data, which forms an alternative way to capture and consider the temporal dynamics of the transportation system in the proposed model. However, it will rule out the certain group of population not using smart phones, and it may bring the privacy issues since it could be possible to identify individuals from even anonymized datasets (De Montjoye et al., 2013).

**Acknowledgements**
We thank the editor and the three anonymous referees for their valuable comments that greatly improved the manuscript. We also thank Dr. Joseph S. DeSalvo, Professor of Economics, University of South Florida for his generous help with identifying job centers in the Tampa Bay region.

Kain, J. F. (1968). Housing segregation, negro employment, and metropolitan decentralization. *The Quarterly Journal of Economics*, *82*(2), 175-197.

Kobayashi, T., Medina, R. M., & Cova, T. J. (2011). Visualizing diurnal population change in urban areas for emergency management. *The Professional Geographer*, *63*(1), 113-130.

Kwan, M. P. (1998). Space-time and integral measures of individual accessibility: a comparative analysis using a point-based framework. *Geographical Analysis*, *30*(3), 191-216.

Kwan, M. P. (2000). Interactive geovisualization of activity-travel patterns using three-dimensional geographical information systems: a methodological exploration with a large data set. *Transportation Research Part C: Emerging Technologies*, *8*(1-6), 185-203.

Le Vine, S., Lee-Gosselin, M., Sivakumar, A., & Polak, J. (2013). A new concept of accessibility to personal activities: development of theory and application to an empirical study of mobility resource holdings. *Journal of Transport Geography*, *31*, 1-10.

Lee, W. K., Sohn, S. Y., & Heo, J. (2018). Utilizing mobile phone-based floating population data to measure the spatial accessibility to public transit. *Applied Geography*, *92*, 123-130.

Levinson, D. M. (1998). Accessibility and the journey to work. *Journal of Transport Geography*, *6*(1), 11-21.

Lin, T. G., Xia, J. C., Robinson, T. P., Goulias, K. G., Church, R. L., Olaru, D., Tapin, J., & Han, R. (2014). Spatial analysis of access to and accessibility surrounding train stations: a case study of accessibility for the elderly in Perth, Western Australia. *Journal of Transport Geography*, *39*, 111-120.

Luo, W., & Wang, F. (2003). Measures of spatial accessibility to health care in a GIS environment: synthesis and a case study in the Chicago region. *Environment and Planning B: Planning and Design*, *30*(6), 865-884.

Matas, A., Raymond, J. L., & Roig, J. L. (2010). Job accessibility and female employment probability: the cases of Barcelona and Madrid. *Urban Studies*, *47*(4), 769-787.

Mennis, J., & Hultgren, T. (2006). Intelligent dasymetric mapping and its application to areal interpolation. *Cartography and Geographic Information Science*, *33*(3), 179-194.

Merlin, L. A., & Hu, L. (2017). Does competition matter in measures of job accessibility? Explaining employment in Los Angeles. *Journal of Transport Geography*, *64*, 77-88.
25